\documentclass{mn2e}

\usepackage{times}
\usepackage{bm}
\usepackage{graphics}
\usepackage{astrobib}

\newcommand{\be}{\begin{equation}}
\newcommand{\ee}{\end{equation}}
\newcommand{\beqa}{\begin{eqnarray}}
\newcommand{\eeqa}{\end{eqnarray}}
\def\d{{\mathrm{d}}}

\title[GRBs as dark energy-matter probes]
{Gamma-ray bursts as dark energy-matter probes in the context of
the generalized Chaplygin gas model.}
\author[O. Bertolami \& P.T. Silva]
{O.~Bertolami\thanks{E-mail: orfeu@cosmos.ist.utl.pt}
and P.T.~Silva\thanks{E-mail:paptms@ist.utl.pt}\\
Instituto Superior T\'ecnico, Departamento de F\'\i sica, Avenida Rovisco 
Pais, 1, 1049-001, Lisboa, Portugal}

\date{Accepted .
      Received ;
      }

\pagerange{\pageref{firstpage}-\pageref{lastpage}}
\pubyear{2005}

\begin{document}

\maketitle
\label{firstpage}

%%%%%%%%%%%%%%%%%%%%%%%%%%%%%%%%%%%%%%%%%%%%%%%%%%%%%%%%%%%%%%%%%%%%%%%%%%%%
\begin{abstract}

\noindent
In this paper we consider the use of Gamma Ray Bursts (GRBs)  as distance markers
to study the unification of dark energy and dark matter in the context of 
the so-called Generalized Chaplygin Gas (GCG) model. We consider
that the GRB luminosity may be estimated from its variability, time-lag, and also use
the so-called Ghirlanda relation. We evaluate the improvements one may expect once more
GRBs and their redshift become available. We show that
although GRBs allow for extending the Hubble diagram to higher redshifts, its
use as a dark energy probe is limited when compared to SNe Ia.
We find that the information from GRBs can provide some bounds on the amount
of dark matter and dark energy independently of the equation of state.
This is particularly evident for XCDM-type models, which are, for 
low-redshifts ($z\leq2$), degenerate with the GCG.

\end{abstract}

%%%%%%%%%%%%%%%%%%%%%%%%%%%%%%%%%%%%%%%%%%%%%%%%%%%%%%%%%%%%%%%%%%%%%%%%%%%%

%%%%%%%%%%%%%%%%%%%%%%%%%%%%%%%%%%%%%%%%%%
%%%%%%%%%%%%%%%%%%%%%%%%%%%%%%%%%%%%%%%%%%
\section{Introduction}

The GCG model \cite{kamenshchik2001,bento2002} is an interesting alternative 
to more conventional approaches for explaining the observed accelerated
expansion of the Universe such as a
cosmological constant (see e.g. \citeNP{bento1999}; \citeNP{bento2001}) or quintessence
(\citeNP{ratra88a}, \citeyearNP{ratra88b};
\citeNP{wetterich88,caldwell98,ferreira98,zlatev99,binetruy99,kim99,uzan99,chiba99,amendola99}; \citeNP{albrecht00,fujii00,bertolami00,sen01a,sen01b,bento02}).
It is worth remarking that quintessence is related to the idea that the
cosmological term could 
evolve \cite{bronstein1933,bertolami1986a,bertolami1986b,ozer1987}
and with attempts to solve the cosmological constant problem. 

In the GCG approach one considers an exotic equation of state 
to describe the behaviour of the background fluid:

\be
p_{ch} = - {A \over \rho_{ch}^{\alpha}} ~~,
\label{eqstate}
\ee
where $A$ and $\alpha$ are positive constants. The case $\alpha=1$
corresponds to the Chaplygin gas. In most phenomenological studies 
the range $0 < \alpha \le 1$ is considered. Within the
framework of Friedmann-Robertson-Walker cosmology, this equation of state
leads, after being inserted into the relativistic energy conservation equation, to
an evolution of the energy density as \shortcite{bento2002}

\be
\rho_{ch}=  \left[A + {B \over a^{3 (1 + \alpha)}}\right]^{1 \over 1 +
\alpha}~~,
\label{rhoch}
\ee
where $a$ is the scale-factor of the Universe and $B$ a positive
integration constant. From this result, one can understand a striking property of the GCG: 
at early times the energy density behaves as matter while at late times it behaves
like a cosmological constant. This behaviour suggests the interpretation of 
the GCG model as an entangled mixture of dark matter and dark energy.

This model has several attractive features. From a theoretical point of view, 
the pure Chaplygin model ($\alpha=1$) equation of state can be
obtained  from the Nambu-Goto action for {\it d}-branes moving in a
$(d+2)$-dimensional spacetime in the light cone
parameterization \cite{bordemann1993}. It is also the only fluid which admits
a supersymmetric generalization \cite{jackiw2000}, and also
appeared in the study of the stabilization of branes in bulks
with a black hole geometry \cite{kamenshchik2000}. The Chaplygin gas may be viewed as a 
quintessence field with a suitable potential \shortcite{kamenshchik2001},
or as an effect arising from the embedding of a $(3+1)$-dimensional
brane in a $(4+1)$-dimensional bulk \cite{bilic2002}. The generalized
Chaplygin gas also has a connection with brane theories \shortcite{bento2002}.
The model can be yet viewed as the simplest model within the family of
tachyon cosmological models \cite{frolov2002}.

The GCG model has also been successfully confronted with different classes of 
phenomenological tests: high precision Cosmic Microwave Background
Radiation data \cite{bento2003a,bento2003b,bento2003c,caturan2003,amendola2003},
supernova data \cite{fabris2002a,dev2003,gorini2003,makler2003,alcaniz2003,bertolami2004,bento2005},
and gravitational lensing \cite{silva2003,dev2004}. More recently, 
it has been shown using the latest supernova data
\cite{tonry2003,barris2004,riess2004}, 
that the GCG model is degenerate with a dark energy model with 
a phantom-like equation of state \shortcite{bertolami2004,bento2005}. Furthermore, 
it can be shown  that this 
does not involve any violation of  the dominant energy condition and hence does
not lead to the big rip singularity in future \shortcite{bertolami2004}. 
It is a feature of GCG model, that it can mimic a phantom-like equation of state, 
but without any kind of pathologies 
as asymptotically the GCG approaches to a well-behaved de-Sitter
universe. Structure formation in the context of the Chaplygin gas and the GCG
was originally examined  in \shortciteN{bento2002}, \shortciteN{bilic2002}
and \citeN{fabris2002b}.
The results of the various phenomenological tests on the GCG model are summarized
in \citeN{bertolami2004a} and \citeN{bertolami2005},

Subsequently, concerns about such an
unified model were raised in the context of structure formation. Indeed, it 
has been pointed out that one should expect unphysical oscillations or even an
exponential blow-up in the matter power spectrum \cite{sandvik2004},  
given the behaviour of the sound
velocity through the GCG. Although, at  early times, the GCG
behaves like  dark matter and its sound velocity is vanishingly small, at later times
the GCG starts behaving like dark energy
with a substantial negative pressure yielding a large sound velocity
which, in turn, can produce oscillations or blow-up in the power
spectrum. This is a common feature of any unified approach when 
the dark matter and the dark energy components of the fluid are not clearly identified. 
These components are, of course, interacting, as they make part of the same fluid.
It can be shown however that the GCG is a unique mixture of
interacting dark matter and a cosmological constant-like dark energy,
once one excludes the possibility of phantom-type dark energy \cite{bento2004}. 
It can be shown that due to the interaction between the components, there is a flow of energy from
dark matter to dark energy. This energy
transfer is not significant until the  recent past, resulting in a
negligible contribution at the time of
gravitational collapse ($z_c \simeq 10$). Subsequently, 
at about $z \simeq 2$), the interaction starts to
grow yielding a large energy transfer from dark matter to dark
energy, which leads to the dominance of the latter at present. Actually, it is shown that
the epoch of dark energy dominance occurs when dark
matter perturbations start deviating from its linear behaviour
and that the Newtonian equations for small scale
perturbations for dark matter do not involve any mode-dependent
term. Thus, neither oscillations nor blow-up in the power spectrum do develop. 

In this paper we study yet another cosmological test and its possible use to study the GCG. 
\citeN{schaefer2002} suggested that Gamma Ray Bursts (GRBs) may be used to extend the
Hubble diagram to redshifts as high as $z\sim 5$. For `ordinary' dark energy, such high
redshifts are not very interesting since at those epochs the Universe is dominated by 
dark matter, and thus it is less sensitive to the nature of dark energy. 

For the GCG however, the GRB test might be relevant since
it unifies dark energy and dark matter into one single fluid. Therefore, within the framework of the
GCG model, the dark matter domination period actually depends of the nature of the
dark energy component that kicks in at later times, and one can expect that
the study of the matter dominated era will bring some insight on some properties of GCG models.

This paper is organized as follows; in section 2 we explain our method of using the
time-lag/luminosity and variability/luminosity correlations to constrain cosmological models.
In section 3 we present and comment the results obtain from  this method.
In section 4 we consider a more precise correlation found by \citeN{ghirlanda2004a}, and
study its consequences. In section 5 we consider whether it is possible to use the extended redshift
range of GRBs to break the degeneracy between the GCG and the XCDM (cold dark matter plus a
dark energy component with a constant equation of state). Finally in section 6 we discuss
our results and present our conclusions.

\section{Method.}

%%%%%%%%%%%%%%%%%%%%%%%%%%%%%%%%%%%%%%%%%%
%%%%%%%%%%%%%%%%%%%%%%%%%%%%%%%%%%%%%%%%%%
\subsection{Overview.}

The starting point of our study is the proposed correlation between time
lags in GRBs spectra and the isotropic equivalent luminosity
\cite{norris2000}, and the correlation between GRB variability and isotropic equivalent 
luminosity \cite{reichart2001}. The time lag, denoted by $\tau_{lag}$,
measures the time offset between high and low energy GRB photons
that arrive on Earth. The variability, $V$,
is easily defined in qualitative terms as a measurement of the
``spikiness'' or complexity of the GRB light curve. The isotropic 
equivalent luminosity is the inferred luminosity of a GRB if all
its energy is radiated isotropically.
That is, if P is the peak flux of a burst in units of photons cm$^{-2}$s$^{-1}$
between observer frame energies $E_l$ and $E_u$, the isotropic
equivalent peak photon luminosity of the burst in $erg~s^{-1}$ between
source frame energies $300~keV$ and $2000~keV$ is given by
\be
\label{Liso}
L_{iso}=4\pi r^2(z) P {\int_{300}^{2000}E N[E/(1+z)]dE\over
\int_{E_l}^{E_u}N(E)dE}
\ee
where $N(E)$ is the observer frame spectral shape, usually
parameterized by a Band function \cite{band1993}, and
$r(z)$ is the comoving distance to a burst at redshift $z$.

The possible use of this relation to expand the Hubble diagram to
higher redshifts was first discussed in \citeN{schaefer2002}. One
limitation of the employed method is the cosmological distances of GRBs,
which affect the ability of performing a proper calibration of their
distance independently of the background cosmology. One has to either fix
a cosmological model and find a calibration that depends on the
assumed cosmological model, or fit the data to both
calibration \emph{and} cosmological parameters. This degrades 
precision since there are more free parameters for the same
number of data points.

One way around this was proposed by \citeN{takahashi2003}. Let us assume
that one measures the luminosity distance up to $z=z_{max}$, with say
$z_{max}=1.5$. This is, for instance, very likely to be possible with the
SNAP\footnote{www.snap.gov} mission. This means
that one would have an estimate of the absolute isotropic luminosity
that is independent of the calibration and of the cosmological model.
One may then use these estimates to calibrate the $(\tau_{lag},L_{iso})$
and $(V,L_{iso})$ relations without assuming a background cosmological model.

The major strength of GRBs as cosmological probes is that they can be found
at very high redshifts \cite{lamb2000}. One may then use
high-redshift GRBs data to probe the cosmology. This is performed by using
the $(\tau_{lag},L_{iso})$ and $(V,L_{iso})$ relations to estimate
the luminosity distance at higher redshifts. A possible method would
consist in using GRBs with $z<1.5$ together with the luminosity distance
estimates from supernovae to calibrate the luminosity estimator, and then
use this calibration to find the luminosity distance of GRBs with $z>1.5$.
It should be noted that this method aims to study the luminosity distance
at the range $1.5<z<5$. Later it will be shown that adding information
obtained in the range $z<1.5$ is actually crucial to the study of dark energy models.

With this information, one may estimate the luminosity distance at
high redshifts, and place constraints on the cosmological 
parameters via a standard $\chi^2$ minimization procedure.

The aim of this paper is to use this method to constrain the GCG unification model
of dark energy and dark matter. Even though it is shown that the
optimum redshift range for studying dark energy is around $z<2$ 
\cite{huterer2000}, as already remarked, since the GCG also describes dark matter,
a higher redshift range might be relevant for a better understanding of the model.

Our study will be performed in three steps.
First we build a realistic mock distribution of GRBs in redshift
and isotropic luminosity space (section 2.2). 
Second, we test the calibration procedure to find what improvements
might be achieved in the future. To do this we shall consider a fiducial set
of calibration parameters to generate a mock set of time lags
and variabilities for each GRB. We shall then perform a $\chi^2$ fit to 
this mock data to study the calibration precision.
The last step consists in employing this method to probe
GCG models. This will be done in a fashion similar to what was already performed with
SNe Ia \cite{goliath2001,weller2002,pietro2003,silva2003}. A fiducial cosmological model
is assumed, and regions of constant $\chi^2$ will be plotted around the
fiducial set of parameters.

%%%%%%%%%%%%%%%%%%%%%%%%%%%%%%%%%%%%%%%%%%%%%%%%
%%%%%%%%%%%%%%%%%%%%%%%%%%%%%%%%%%%%%%%%%%%%%%%%
\subsection{Generating a GRB mock population.}

We describe here the process of population generation.
We simulate several data sets, which differ only in size. We consider
three sample sizes. The first sample is composed of 90 GRBs, and is
consistent with the expected number of GRBs with measured redshift $z>1.5$
that is what the SWIFT satellite is expected to detect in its three year mission
\cite{schaefer2002}. The  second and third samples are larger and contain 500 and 1000 GRBs,
respectively, and serve as best case scenarios to test whether further data
might improve the results.

The GRBs in each sample are distributed in redshift and luminosity according
to the GRB rate history and luminosity function based on the model
of the star formation 2 from \citeN{porciani2000}. The energy spectrum of the 
GRB is assumed to be a single power law with index -2.5. The flux limit of the SWIFT
satellite, $f>0.04$ photons cm$^{-2}$s$^{-1}$ is applied to check whether the GRB
can be detected. The observed magnitude is calculated assuming a flat
$\Lambda$CDM cosmological model, with $\Omega_\Lambda=0.7$ and $H_0=70~km~s^{-1}
~Mpc^{-1}$.

\begin{figure}
\begin{center}
{\resizebox{0.45\textwidth}{!}{\includegraphics{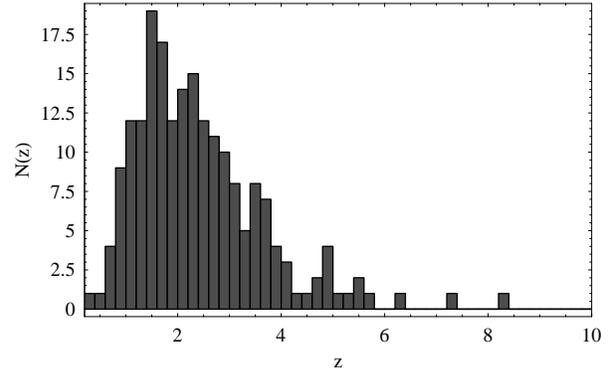}}}
\caption{The redshift distribution of a GRB sample population.}
\label{figure1}
\end{center}
\end{figure}

We show an example of a generated sample in Figure \ref{figure1}.

%%%%%%%%%%%%%%%%%%%%%%%%%%%%%%%%%%%%%%%%%%%%%%%%%%%%%%%%%%%%%%%%%%%%%%%%%%%%%%%
%%%%%%%%%%%%%%%%%%%%%%%%%%%%%%%%%%%%%%%%%%%%%%%%%%%%%%%%%%%%%%%%%%%%%%%%%%%%%%

%%%%%%%%%%%%%%%%%%%%%%%%%%%%%%%%%%%%%%%%%%
%%%%%%%%%%%%%%%%%%%%%%%%%%%%%%%%%%%%%%%%%%
\subsection{Calibration procedure.}

\begin{table}
\begin{center}
\begin{tabular}{lr}
\hline
\hline
Parameter & Value\\
\hline
$B_v$&$10^{55.32}$\\
$\beta_v$&$1.57$\\
$\sigma_v$&$0.20$\\
$B_\tau$&$10^{50.03}$\\
$\beta_\tau$&$-1.27$\\
$\sigma_\tau$&$0.35$\\
\hline
\end{tabular}
\caption{Values used in the calibration test.}
\end{center}
\end{table}

\citeN{reichart2001} proposed a relation between
the variability and isotropic equivalent luminosity of a GRB such that
\be
\label{VL}
L_{iso}=B_vV^{\beta_v},
\ee
while \shortciteN{norris2000} proposed a similar relation between $\tau_{lag}$ and
$L_{iso}$,
\be
\label{TL}
L_{iso}=B_\tau \tau_{lag}^{\beta_v}.
\ee

As already mentioned, the first step in testing the calibration procedure
consists in establishing a fiducial model. \citeN{schaefer2002} used the nine
GRBs with available redshifts at the time to calibrate these relations,
yielding:
\beqa
\beta_v=1.57~;~~~~~~~~B_v=10^{50.03}~;\\
\beta_\tau=-1.27~;~~~~~~~~B_\tau=10^{55.32}.
\eeqa
We shall assume these values as our fiducial model, i.e. we suppose
that the calibration relations are faithful, and that they are described
by this set of parameters. For each GRB of the mock luminosity distribution,
 generated as we explained above, we compute the corresponding
time lag, $\tau_{lag}$, and variability, $V$, through
\beqa
\label{LogTau}
\log \tau_{lag}=\log B_\tau~+~{1\over\beta_\tau}\log L+Random(\sigma_\tau),\\
\label{LogV}
\log V=\log B_v~+~{1\over\beta_v}\log L+Random(\sigma_v),
\eeqa
where the $Random(\sigma)$ term is a pseudo random number drawn from a normal
distribution with zero mean value and variance $\sigma$. That is, we assume that
$\tau_{lag}$ and $V$ have lognormal error distributions, with variance
$\sigma_\tau$ and $\sigma_v$, respectively.

The values we use for $\sigma_v$ and $\sigma_\tau$ are based on \citeN{schaefer2002}
and are essentially dominated by an intrinsic (statistical) error. The
used fiducial values are shown in Table 1.

In the real situation, we would assume that the luminosity distance
at redshifts smaller than $z=1.5$ has been measured independently by SNe Ia
experiments, such as SNAP. Knowledge of the peak flux observed on Earth, and
the spectral shape $N(E)$ allow obtaining  the corresponding equivalent isotropic
luminosity from Eq. (\ref{Liso}).

We do not consider errors in this estimate of $L_{iso}$, but it should be
noted that it is expected that the luminosity distance uncertainty will be
around  $\sigma_{\log d_L}\sim 0.01$ for future experiments such as
SNAP \cite{goliath2001,weller2002}. This translates into $\sigma_{\log L}\sim 0.02$, a
much smaller error than the intrinsic scatter of GRBs. Thus, this source of uncertainty
should not influence our conclusions.

Notice that we have used low-redshift GRBs from the generated sample
to calibrate the relations Eqs. (\ref{VL}) and (\ref{TL}) through a standard
$\chi^2$ fitting procedure. That is, our procedure consisted in:
\begin{itemize}
\item{Generating a GRB sample.}
\item{For each GRB in the sample, assigning a time lag and variability via
Eqs. (\ref{LogTau}) and  (\ref{LogV}).}
\item{Fitting the calibration relations Eqs. (\ref{VL}) and (\ref{TL}) for the 
low-z GRBs from the mock sample.}
\item{Repeating the steps for different size samples, and study the 
precision of the attained fits.}
\end{itemize}

Our results are shown in Table 2. The smaller sample size
corresponds to what is expected to be found by the SWIFT satellite in 
its three year campaign, the second sample size corresponds to the
number of low-z GRBs that should be found if  there where about $500$
GRBs in total, while the third sample corresponds to about $1000$ GRBs 
in total.

\begin{table}
\begin{center}
\begin{tabular}{cccc}
\hline
\hline
$N_{GRB}$ used in calibration & $\sigma_{\beta_\tau}$ & $\sigma_{\beta_V}$
& $\sigma_\mu$(mag.)\\ 
\hline
$40$ & $0.053$ & $0.021$ & $0.68$ \\
$100$ & $0.033$ & $0.014$ & $0.66$ \\
$200$ & $0.023$ & $0.010$ & $0.66$ \\
\hline
\end{tabular}
\caption{Results from calibration test. The last column refers to the
uncertainty of the distance modulus determination using our method.}
\end{center}
\end{table}

%%%%%%%%%%%%%%%%%%%%%%%%%%%%%%%%%%%%%%%%%%
%%%%%%%%%%%%%%%%%%%%%%%%%%%%%%%%%%%%%%%%%%
\subsection{ Estimating $L_{iso}$ and the $\chi^2$ test.}

For the remaining GRBs that were not used in the calibration procedure,
the calibrated relations will be used to find $L_{iso}$ from the values of
$\tau$ and $V$. Thus we obtain two estimates:
\beqa
\label{LV}
\log L_V=&\log B_v +\beta_v\log V,\\
\label{LT}
\log L_\tau=&\log B_\tau +\beta_\tau\log\tau_{lag}.
\eeqa
These two estimates are combined as weighted averages to produce one 
estimate of $L_{iso}$,
\be
\log L_{iso}={1 \over 2} \left( {1 \over \sigma_{\log L_V}}\log L_V+
{1 \over \sigma_{\log L_\tau}}\log L_\tau \right).
\ee

The next step consists in defining the $\chi^2$ function,
\be
\chi^2(\bm{p})=\sum_{i}^{N_{GRB}}\left({m_i-M_i-25-5
\log d_L(z_i,\bm{p})\over\sigma_\mu}\right)^2,
\ee
where $m_i$ is the observed magnitude, $M_i$ is the absolute magnitude
estimated from Eqs. (\ref{LV}) and (\ref{LT}), and $d_L(z_i,\bm{p})$ is the
standard luminosity distance as function of redshift $z$ and the
cosmological parameters $\bm{p}$. The denominator $\sigma_\mu$ is the
uncertainty in the determination of the distance modulus, $m_i-M_i$.
This uncertainty was calculated using Gaussian error propagation, 
and assuming that there were no correlation terms  to 
be taken into account. The values of $\sigma_\mu$ we have found are shown in
Table 2. We have to minimize this function and draw the $\chi^2$ contours in
order to find the confidence regions.

In here we follow the method developed for the study of SNe Ia 
\cite{goliath2001,weller2002,pietro2003,silva2003}. We use the log-likelihood function
$\chi^2$ to build confidence regions in the parameter space. In order to
perform this one
chooses a fiducial model, denoted by the parameter vector $\bm{p_{fid}}$,
and then the log-likelihood functions $\chi^2$ are calculated based on
hypothetical magnitude measurements at the various redshifts. The $\chi^2$
function is then given by
\be
\chi^2(\bm{p})= \sum_{i}^{N_{GRB}}\left({5\log d_L(z_i,\bm{p_{fid}})-5\log d_L(z_i,\bm{p})\over\sigma_\mu}\right)^2~~.
\ee
Of course, there is no need to minimize the $\chi^2$ 
function since its minimum will correspond to $\bm{p}=\bm{p_{fid}}$,
thus we only need to find the $\chi^2$ contours corresponding 
to the desired confidence level.

%%%%%%%%%%%%%%%%%%%%%%%%%%%%%%%%%%%%%%%%%%
%%%%%%%%%%%%%%%%%%%%%%%%%%%%%%%%%%%%%%%%%%
\subsection{The models.}

The GCG model smoothly interpolates between
a dark matter dominated time in the past, to an accelerated de Sitter phase
in the future. Thus, this is a setting that on large scales agrees
with the observed expansion history of the Universe
\shortcite{kamenshchik2001,bento2002}. The GCG density may be written
as a function of redshift as
\be
\rho_{ch}(z)=\rho_{ch,0}\left[A_s+(1-A_s)(1+z)^{3(1+\alpha)}
\right]^{1/(1+\alpha)}
\ee
where $\rho_{ch,0}$ is the present day density of the GCG, and
\be
A_s\equiv A \rho_{ch,0}^{-(1+\alpha)};~~~~~~(A+B)^{1/(1+\alpha)}=\rho_{ch,0}~~,
\ee
where $B$ is the integration constant that appears in Eq. (\ref{rhoch}).
For $z\gg0$ we have a matter dominated Universe,
\be
\label{ChapHighz}
\rho_{ch}(z\gg0)=(1-A_s)^{1/(1+\alpha)}(1+z)^3,
\ee
while in the far future, $z=-1$ the GCG behaves as a vacuum dominated Universe.

The GCG unifies dark matter and dark energy, but does not take into account the
presence of baryons and radiation.
The baryonic component has to be considered when studying the implications
of the GCG model with regards to observational tests, but for 
geometric tests, such as the magnitude-redshift relation we consider here,
the effect of these components is negligible. Therefore, throughout the
paper we disregard the presence of baryons and radiation. This does not affect 
any of our conclusions.

For the purpose of comparison we choose the so-called XCDM model, which consists
of two components, cold dark matter and some
form of dark energy which has a constant negative equation of state, 
$w=p/\rho$. We shall use this parameterization to test GRBs as a probe of an unspecified
dark energy component. Although this parameterization is not suitable for general
dark energy models, since in most cases the equation of state changes with time,
it is adequate to test the cosmological constant model, $w=-1$, against other models.
If one finds that $w=-1$ is disfavoured by the data, then there is a strong indication
that the dark energy component is more complex than expected.

For many years cosmologists were mainly concerned with  models to which $-1\leq w<-2/3$, however,
more recently mounting evidence that $w<-1$, the so-called phantom dark energy models
\cite{caldwell2002,carrol2003}, is being encountered. In many models there are several
theoretical reasons not to consider $w<-1$, most notably that this would
lead to a breakdown of the weak, the dominant and the strong energy conditions
($\rho\geq0$ and $\rho+p\geq0$, $\rho\geq|p|$, and $\rho+3p\geq0$, respectively).
The weak and strong energy conditions are ones to hold in proving well known
singularity theorems, while the dominant energy condition
guarantees the stability of a component, such that there is no creation of
energy-moment from nothing \cite{hawking1973}.

Another feature of some phantom models is that they exhibit a future blow up of the
scale factor in a finite time, often referred to as the `Big-Rip'. 
Since such a Universe has a finite lifetime, it has been 
argued that these phantom models solve the coincidence problem, that is, they explain
why dark matter and dark energy densities are of the same order at the present time.
For a $\Lambda$CDM Universe, a long matter dominated period is followed by
a rather quick transition to a phase such as the one the Universe is now, where
matter and dark energy have approximate densities. This phase is then followed by
an eternal vacuum dominated exponential expansion. Within this framework, 
the probability of finding ourselves in this intermediate and
temporary epoch is very small. For phantom models on the other hand, the accelerated
expansion is not eternal.
Since the Universe has a finite lifetime, the probability of living in the epoch of
matter-energy approximate equality is larger than that of $\Lambda$CDM models, thus, 
the cosmic coincidence is not as unlikely as for $\Lambda$CDM models.
However, these models fail to explain why the dark energy component did not start
to dominate the evolution at a earlier time, namely, why dark energy started dominating
the cosmic evolution only after the large scale structure had time to evolve deep into the
non-linear regime, even though in the context of the CGC this is 
exactly what is found \shortcite{bento2004}.

Furthermore, besides these theoretical features, the latest SNe Ia data does seem to favour a
phantom energy component. Thus, in what follows, we shall consider a phantom model for comparison
with the GCG model. Moreover, as shown in \shortciteN{bertolami2004} and \shortciteN{bento2005}, for
$z<2$, the GCG model is degenerate with phantom models with suitable parameters.

As mentioned, we consider two different cosmological models. The flat GCG model,
which unifies dark energy and dark matter into a single
component, and the XCDM, which parameterizes dark energy in terms of a constant equation
of state, $w=p/\rho$. We consider that the Universe is flat in both models. The GCG
model is then described by the exponent $\alpha$, and the quantity $A_s$, while the
XCDM is described by the parameters $w$ and $\Omega_m$, the non-relativistic dark matter
density relative to the critical one. We considered two fiducial models. Model I
 assumes that the Universe is described by a GCG model with $1-A_s=0.3$ and $\alpha=1$.
Model II assumes the corresponding degenerate Universe, described by a phantom XCDM model,
with $w=-1.4$ and $\Omega_m=0.45$.

%%%%%%%%%%%%%%%%%%%%%%%%%%%%%%%%%%%%%%%%%%
%%%%%%%%%%%%%%%%%%%%%%%%%%%%%%%%%%%%%%%%%%
\section{Results.}

\begin{figure}
\begin{center}
{\resizebox{0.45\textwidth}{!}{\includegraphics{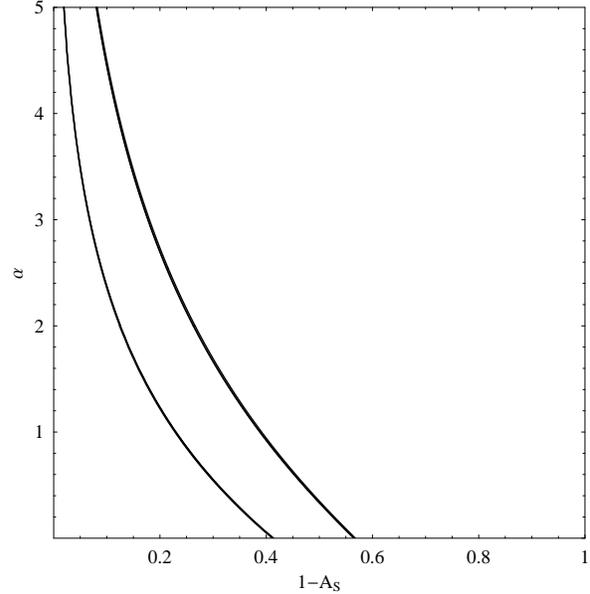}}}
\caption{Effects of the improvement in calibration of GRBs for CGC models.
The three sets of curves are indistinguishable.}
\label{figure2}
\end{center}
\end{figure}

\begin{figure}
\begin{center}
{\resizebox{0.45\textwidth}{!}{\includegraphics{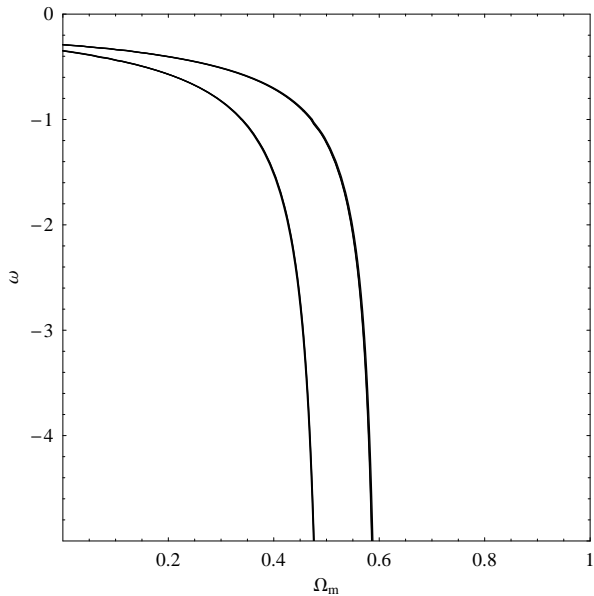}}}
\caption{Effects of the improvement in calibration of GRBs for XCDM models.
The three sets of curves are indistinguishable.}
\label{figure3}
\end{center}
\end{figure}

We first examined the effect that an improvement in the calibration might have in 
constraining the models. To that end we consider that the high-redshift
GRBs sample is comprised by 90 GRBs with redshifts $z>1.5$, and use the
values shown in Table 2 to calculate the uncertainty in the 
determination of $L_{iso}$. The effect that an improvement in calibration
has in the quality of the cosmological tests is shown in Figures \ref{figure2}
and \ref{figure3}.

%%%%%%%%%%%%%%%%%%%%%%%%%%%%%%%%%%%%%%%%%%%%%%%%%%%%%%%%%%%%%%%%%%%%%%%%%%%%%%%
%%%%%%%%%%%%%%%%%%%%%%%%%%%%%%%%%%%%%%%%%%%%%%%%%%%%%%%%%%%%%%%%%%%%%%%%%%%%%%

%%%%%%%%%%%%%%%%%%%%%%%%%%%%%%%%%%%%%%%%%%%%%%%%%%%%%%%%%%%%%%%%%%%%%%%%%%%%%%
%%%%%%%%%%%%%%%%%%%%%%%%%%%%%%%%%%%%%%%%%%%%%%%%%%%%%%%%%%%%%%%%%%%%%%%%%%%%%%

\begin{figure}
\begin{center}
{\resizebox{0.45\textwidth}{!}{\includegraphics{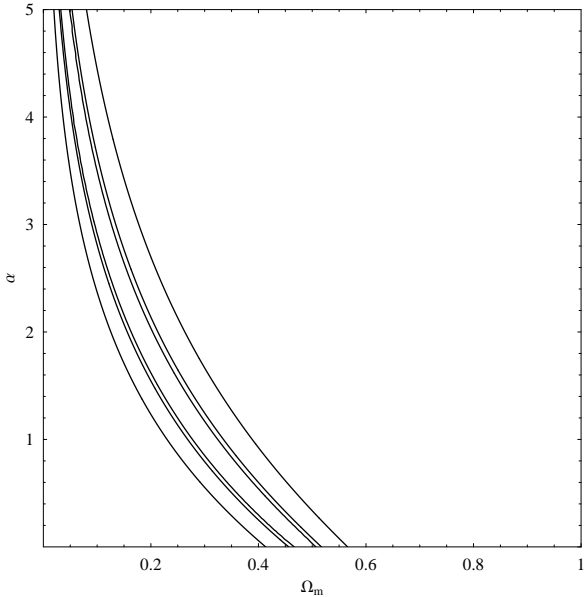}}}
\caption{Effects of the increase in the number of GRBs for CGC models. The curves
correspond, from the outer to the inner ones, to 90, 500 and 1000 high-redshift GRBs}
\label{figure4}
\end{center}
\end{figure}

\begin{figure}
\begin{center}
{\resizebox{0.45\textwidth}{!}{\includegraphics{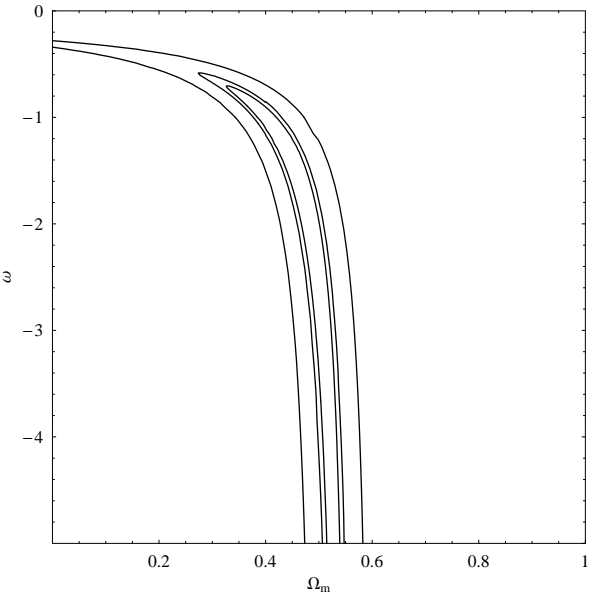}}}
\caption{Effects of the increase in the number of GRBs for XCDM models. The
curves correspond, from the outer to the inner curves, to 90, 500 and 1000
high-redshift GRBs.}
\label{figure5}
\end{center}
\end{figure}

We then tested the effect of an increase in the number of high-redshift GRBs
in the test. We assume for the purpose of calibration that 100 low-redshift
GRBs are known, and use the corresponding calibration precision shown in
Table 2. We next considered larger samples of GRBs. Samples with 90, 500
and 1000 high-redshift GRBs are examined, and the respective confidence
regions are exhibited. These results are shown in Figures
\ref{figure4} and \ref{figure5}

%%%%%%%%%%%%%%%%%%%%%%%%%%%%%%%%%%%%%%%%%%%%%%%%%%%%%%%%%%%%%%%%%%%%%%%%%%%%%%
%%%%%%%%%%%%%%%%%%%%%%%%%%%%%%%%%%%%%%%%%%%%%%%%%%%%%%%%%%%%%%%%%%%%%%%%%%%%%%

%%%%%%%%%%%%%%%%%%%%%%%%%%%%%%%%%%%%%%%%%%%%%%%%%%%%%%%%%%%%%%%%%%%%%%%%%%%%%%
%%%%%%%%%%%%%%%%%%%%%%%%%%%%%%%%%%%%%%%%%%%%%%%%%%%%%%%%%%%%%%%%%%%%%%%%%%%%%%

\begin{figure}
\begin{center}
{\resizebox{0.45\textwidth}{!}{\includegraphics{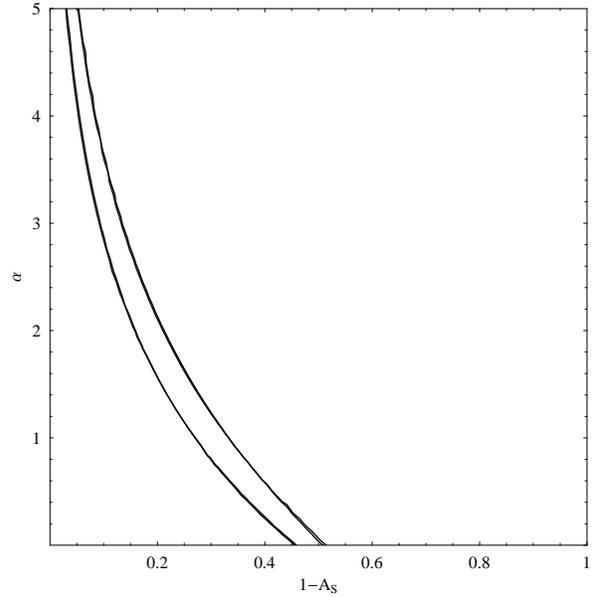}}}
\caption{Effect of adding 100 low-redshift GRBs to a sample of 500 high-redshift 
GRBs for CGC models. Both sets of curves are indistinguishable.}
\label{figure6}
\end{center}
\end{figure}

\begin{figure}
\begin{center}
{\resizebox{0.45\textwidth}{!}{\includegraphics{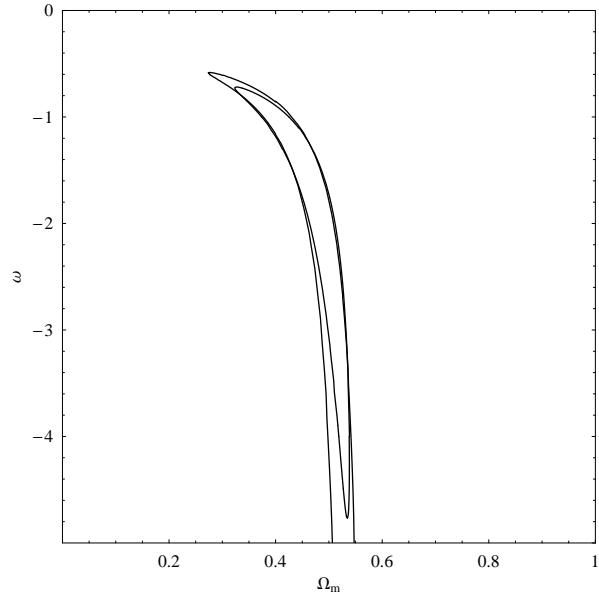}}}
\caption{Effect of adding 100 low-redshift GRBs to a sample of 500 high-redshift 
GRBs for XCDM models. The outer curve corresponds to 500 high-redshift GRBs, while
the inner one corresponds to the 100+500 GRB case.}
\label{figure7}
\end{center}
\end{figure}

Finally we study the effect of the distribution of GRBs in redshift space. In
order to perform this analysis we added 100 GRBs with $z<1.5$ to a sample of
400 high-redshift GRBs, and found the corresponding confidence regions. The
results are shown in  Figures \ref{figure6} and \ref{figure7}.

%%%%%%%%%%%%%%%%%%%%%%%%%%%%%%%%%%%%%%%%%%%%%%%%%%%%%%%%%%%%%%%%%%%%%%%%%%%%%%
%%%%%%%%%%%%%%%%%%%%%%%%%%%%%%%%%%%%%%%%%%%%%%%%%%%%%%%%%%%%%%%%%%%%%%%%%%%%%%

%%%%%%%%%%%%%%%%%%%%%%%%%%%%%%%%%%%%%%%%%%%%%%%%%%%%%%%%%%%%%%%%%%%%%%%%%%%%%%
%%%%%%%%%%%%%%%%%%%%%%%%%%%%%%%%%%%%%%%%%%%%%%%%%%%%%%%%%%%%%%%%%%%%%%%%%%%%%%

 The first conclusion we can draw is that GRBs are not quite suitable as stand-alone
probes of dark energy.  This is seen from the lack of impact that
an increase in the redshift range has on the results in terms
of precision and discriminating power. 

The main source of uncertainty comes from the intrinsic statistical scatter
of the GRB population. As may be seen from Figures \ref{figure2} and 
\ref{figure3}, calibrating the $(\tau,L_{iso})$ and $(V,L_{iso})$ relations
with more than about $30$ low-redshift GRBs does not substantially affect
the results, however an increase in the size of the higher redshift GRB
population will improve the quality of the constraints that can be imposed.
This can be seen in Figures \ref{figure4} and \ref{figure5}. Using more precise
luminosity estimators, such as the one we will use in the following section,
might improve the outcome, even though better results
can be achieved by probing a greater sample of $z<1.5$ redshift sources.
This is evident for XCDM models, as shown in Figure \ref{figure7}. Adding more
$100$ GRBs with $z<1.5$ to a sample of $400$ high-redshift GRBs is more effective
than adding more $500$ GRBs with $z>1.5$. 

Regarding the results for each model, the first conclusion is that the
$XCDM$ model is better constrained than the $CGC$ model. It can be seen that,
no constraint in the $\alpha$ parameter for the GCG model can be imposed, even
though an  upper limit for $A_s$ can be obtained.

As for XCDM models, we point out that there is some potential to probe $\Omega_m$, 
but the prospect of using this test to probe the nature of dark energy
is very limited. A upper value for $w$ can be obtained, however no lower
limit. It should be noted that other tests such as SNe Ia are capable of 
imposing tighter constrains on $\Omega_m$ than the ones we find.

%%%%%%%%%%%%%%%%%%%%%%%%%%%%%%%%%%%%%%%%%%
%%%%%%%%%%%%%%%%%%%%%%%%%%%%%%%%%%%%%%%%%%
\section{The Ghirlanda relation.}

\subsection{Description.}
The $(\tau,L_{iso})$ and $(V,L_{iso})$ relations are notoriously affected
by the large intrinsic scatter of the data set, which greatly hinders their
use as precision cosmological probes. Quite recently,
\shortciteN{ghirlanda2004a} have found a surprisingly tight correlation between the peak
energy of the $\gamma$-ray spectrum, $E_{peak}$ (in the $\nu-\nu F_\nu$ plot), and
the collimation corrected energy emitted in $\gamma$-rays, $E_\gamma$, for long
GRBs. This collimation corrected energy measures the energy release
by the GRB taking into account that the energy is beamed into a jet with 
aperture angle $\theta$. 

Let us denote by $E_{iso}$ the isotropically equivalent energy, inferred
from the isotropic GRB emission.
This source frame ``bolometric''  isotropic energy is found 
integrating the best  fit time-integrated spectrum N(E) [photons cm$^{-2}$ keV$^{-1}$] 
over the energy range 1 keV - 10 MeV,
\be
\label{Eiso}
E_{iso}={ 4\pi d_L^2(z) \over 1+z}S_\gamma k~~~
\textrm{erg}~~,
\ee
where $d_L(z)$ is the GRB luminosity distance, $k$ refers to  the
k-correction \cite{bloom2001} and $S_\gamma$ is the fluence,
\be
S_\gamma=\int_{1}^{10^4}EN(E)dE~~,
\ee
where $E$ is in keV.

If the $\gamma$-ray emission is collimated into a jet with aperture $\theta$,
then the true energy emitted is $E_\gamma=E_{iso}(1-\cos\theta)$. Thus, to convert
$E_{iso}$ into $E_\gamma$ and vice-versa, one needs to know the angle
$\theta$. Under the simplifying assumption  of a constant circum-burst density
medium of number density $n$, a fireball emitting a fraction $\eta_\gamma$ of its kinetic
energy in the prompts $\gamma$-ray phase would show a break in its afterglow
light curve when its bulk Lorentz factor $\Gamma$ becomes of the order of 
$\Gamma \simeq 1/\theta$, with $\theta$ given by \cite{sari1999}:
\be
\label{teta}
\theta=0.161 \left( {t_{jet} \over 1+z}\right)^{3/8} 
\left( {n\eta_\gamma \over E_{iso,52} } \right)^{1/8}
\ee
where $t_{jet}$ is the break time in days, and $E_{iso,52}=E_{iso}/10^{52}$.

The Ghirlanda relation \shortcite{ghirlanda2004a,xu2005b} is then expressed as
\be
{ E_\gamma \over 10^{50}\textrm{erg} } = C\left({ E_{p} \over 100~\textrm{keV}}
\right)^a
\ee
were $a$ and $C$ are dimensionless parameters.

Using this result, together with Eqs. (\ref{Eiso}), (\ref{teta}) and the
definition of $E_\gamma$, one obtains an equation for $d_L(z)$,
\be
C\left( E_{p} \over 100~\textrm{keV} \right)^a=\left(1-\cos\theta\right)\left( E_{iso}
\over 10^{50} \right).
\ee
Note that both $E_{iso}$ and $\theta$ depend on $d_L(x)$, as may be seen
from their definitions. Solving this equation for $d_L(z)$ we find an estimate
of the luminosity distance at redshift $z$ in terms of the observables $E_p$,
$t_{jet}$, $n$, and fluence $S_\gamma$.

The error budget for $d_L$ is then found to be \cite{xu2005b}
\begin{eqnarray}
\label{sigmadl}
 \left( {\sigma_{d_L} \over d_L } \right)^2  & = & {1 \over 4} \left[
 \left( { \sigma_{S_\gamma} \over S_\gamma } \right)^2  +
\left( {\sigma _k  \over k} \right)^2 \right] 
+{1 \over 4}{1 \over (1 - \sqrt   {C_{\theta } } )^2 } \times \nonumber \\
%%%%
 & & \left[ \left( {\sigma _C \over C} \right)^2
    + \left( a{\sigma _{E_p^{obs}} \over E_p^{obs} } \right)^2 +
      \left( a{\sigma _a \over a}\ln {E_p \over 100} \right)^2  \right]+ 
      \nonumber \\
%%%%
 & &\left[ {\left( {3\sigma _{t_j} \over t_j } \right)^2
     + \left( {\sigma_{n_0} \over n_0} \right)^2}+ \left( {\sigma_{\eta _\gamma}
     \over \eta_\gamma  (1 - \eta_\gamma  )} \right)^2 \right] \times \nonumber \\
%%%%
&& {1 \over 4}{C_{\theta }\over {(1 - \sqrt   {C_{\theta } } )^2 }}~~,
\end{eqnarray}
where $C_\theta^2=\theta \sin \theta/(8-8\cos\theta)$.

The distance modulus is given by $\mu_{obs}=5\log d_L/10\textrm{pc}$, thus its
error is $\sigma_\mu=5/\ln 5 (\sigma_dL/d_L)$. We use the error estimates
from \cite{xu2005b}, which are consistent with what is found in the 
literature (\shortciteNP{ghirlanda2004a};
\citeNP{ghirlanda2004b,friedman2005,ghisellini2005,xu2005a,mortsell2005}).

These values yield an uncertainty in $\mu_{obs}$ of order $\sigma_{\mu}\simeq 0.5$, which is just
slightly  smaller than the error bars from using both the $(L_{iso},V)$ and $(L_{iso},\tau_L)$ relations.
As may be observed in Table 3, the smaller intrinsic scatter of the Ghirlanda
relation is balanced by its dependence on poorly constrained quantities.
Improving the calibration will not solve the problem, since
the main sources of uncertainty are due to the determination of the peak energy $E_p$, the 
jet break time $t_{jet}$ and the value of the circum-burst density $n$.

\begin{table}
\begin{center}
\begin{tabular}{cccc}
\hline
\hline
& Error & Contribution to $\left(\sigma_{d_l} \over d_l \right)^2$ 
& Percentage of $\left(\sigma_{d_l} \over d_l \right)^2$\\
\hline
$\left( \sigma_{S_\gamma} \over S_\gamma \right)^2$ & $10 \%$ & $0.0027$ & $5\%$ \\
$\left( \sigma_{k} \over k \right)^2$ & $5 \%$ & $0.0006$ & $1\%$ \\
$\left( \sigma_{C} \over C \right)^2$ & $8 \%$ & $0.0027$ & $5\%$ \\
$\left( \sigma_{a} \over a \right)^2$ & $5 \%$ & $0.0025$ & $5 \%$ \\
$\left( \sigma_{E_{p}} \over E_{p} \right)^2$ & $17 \%$ & $0.0259$ & $52 \%$ \\
$\left( \sigma_{n} \over n \right)^2$ & $50 \%$ & $0.0069$ & $14 \%$ \\
$\left( \sigma_{t_{jet}} \over t_{jet} \right)^2$ & $20 \%$ & $0.0081$ & $16 \%$ \\
\hline
\end{tabular}
\caption{Error budget for the Ghirlanda test. The uncertainty in the circum-burst density is assumed to be $50\%$, while the other values are taken from Xu (2005).}
\end{center}
\end{table}

%%%%%%%%%%%%%%%%%%%%%%%%%%%%%%%%%%%%%%%%%%%%%%%%%%%%%%%%%%
%%%%%%%%%%%%%%%%%%%%%%%%%%%%%%%%%%%%%%%%%%%%%%%%%%%%%%%%%%
\subsection{Results.}

%%%%%%%%%%%%%%%%%%%%%%%%%%%%%%%%%%%%%%%%%%%%%%%%%%%%%%%%%%%%%%%%%%%%%%%%%%%%%%
\begin{figure}
\begin{center}
{\resizebox{0.45\textwidth}{!}{\includegraphics{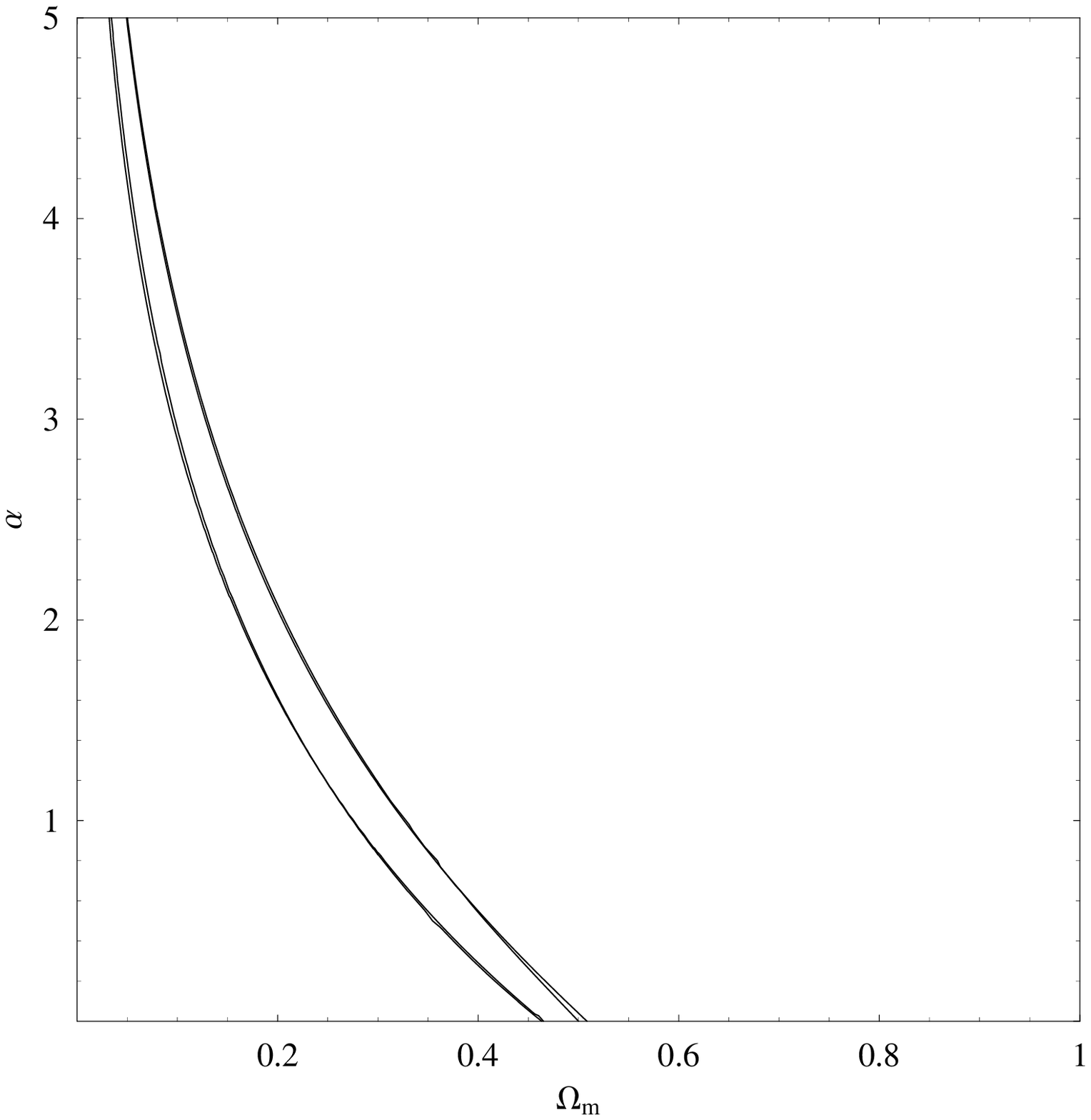}}}
\caption{Same as Figure \ref{figure6}, but using the Ghirlanda relation.}
\label{figure9}
\end{center}
\end{figure}
%%%%%%%%%%%%%%%%%%%%%%%%%%%%%%%%%%%%%%%%%%%%%%%%%%%%%%%%%%%%%%%%%%%%%%%%%%%%%%

%%%%%%%%%%%%%%%%%%%%%%%%%%%%%%%%%%%%%%%%%%%%%%%%%%%%%%%%%%%%%%%%%%%%%%%%%%%%%%
\begin{figure}
\begin{center}
{\resizebox{0.45\textwidth}{!}{\includegraphics{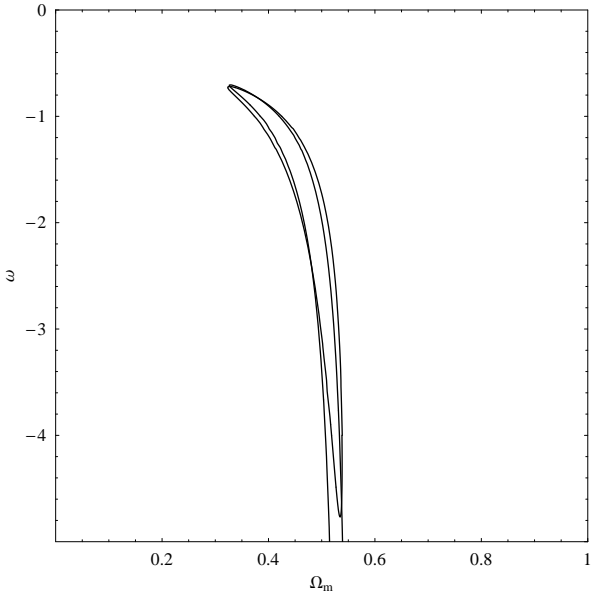}}}
\caption{Same as Figure \ref{figure7}, but using the Ghirlanda relation.}
\label{figure10}
\end{center}
\end{figure}
%%%%%%%%%%%%%%%%%%%%%%%%%%%%%%%%%%%%%%%%%%%%%%%%%%%%%%%%%%%%%%%%%%%%%%%%%%%%%%

In Figures \ref{figure9} and \ref{figure10} we see the improvements that
can be obtained by the use of the Ghirlanda relation.
We have assumed two redshift distributions; one made up of $500$ GRBs with $z>1.5$, 
and the other made up of $100$ GRBs with $z<1.5$ plus $400$ GRBs with $z>1.5$.
We find that the conclusions one can draw remain essentially unchanged. Moreover, we
see that the high-redshifts of GRBs are not quite suitable to adequately study the
GCG model, while it allows for the XCDM model to place a rather useful limit to the
total amount of matter, although not much can be learned with respect to the
dark energy equation of state.

%%%%%%%%%%%%%%%%%%%%%%%%%%%%%%%%%%%%%%%%%%%%%%%%%%%%%%%%%%%
%%%%%%%%%%%%%%%%%%%%%%%%%%%%%%%%%%%%%%%%%%%%%%%%%%%%%%%%%%%

%%%%%%%%%%%%%%%%%%%%%%%%%%%%%%%%%%%%%%%%%%%
%%%%%%%%%%%%%%%%%%%%%%%%%%%%%%%%%%%%%%%%%%%
\section{Degeneracy Between models.}

The CGC and the phantom XCDM models are degenerate at redshifts $z<2$, as 
shown in \shortciteN{bertolami2004} and \citeN{bento2005}. Consider the Taylor
expansion of the luminosity distance as
\beqa
d_L={{c}\over{H_0}}\left\{z+{{1\over2}(1-q_0)}z^2-{{1}\over{6}}
\left(1-q_0-3q_0^2+j_0\right)z^3-\right.
\nonumber\\
-\left[1+{{3}\over{2}}q_0\left(1+q_0\right)+{{5}\over{8}}q_0^3-{{1}\over{2}}j_0
-{{5}\over{12}}q_0 j_0-{{k_0}\over{24}}\right]z^4\nonumber\\
\left.+O(z^5)\right\},
\label{powerDL}
\eeqa
were $q_0$ is the deceleration parameter, related to the second derivative
of the expansion factor, $j_0$ is the so-called ``jerk'' or statefinder parameter
\cite{alam2003,sahni2003}, related to the third derivative of the expansion factor,
and $k_0$ is the so-called ``kerk'' parameter, which is related to the
fourth derivative of the expansion parameter \cite{visser2004,dabrowski2004}, all
evaluated at present. These quantities are defined as
\begin{equation}
q(t) = - {1\over a} \; {\d^2 a\over \d t^2}  \;
\left[ {1\over a} \; {\d a \over  \d t}\right]^{-2};
\end{equation}
\begin{equation}
j(t) = + {1\over a} \; {\d^3 a \over \d t^3}  
\; \left[ {1\over a} \; {\d a \over  \d t}\right]^{-3};
\end{equation}
\begin{equation}
k(t) = + {1\over a} \; {\d^4 a \over \d t^4}  
\; \left[ {1\over a} \; {\d a \over  \d t}\right]^{-4};
\end{equation}
and are related to each other through the relations
\be
q(z)={3\over2}\left( {p(z)\over\rho(z)} +1\right)-1;
\ee
\be
j(z)=q(z)+2q^2(z)+{\d q\over\d z}(z);
\ee
\be
k(z)={\d j\over \d z}(z)-2j(z)-3j(z)q(z).
\ee
were $\rho(z)$ and $p(z)$ refer to the {\it total} density and pressure
of the Universe, respectively. With these expressions one can
write the present day values of the parameters as a function of the
cosmological parameters of the model under examination.

For the $XCDM$ model, we have
\be
q_o^{XCDM}={3 \over 2}\left[1+w(1- \Omega_m - 1)\right]-1,
\ee
\be
\left.{\d q\over \d z}\right|_0^{XCDM}={9\over2} w^2 (1-\Omega_m)
\Omega_m,
\ee
\be
\left.{\d j\over \d z}\right|_0^{XCDM}= -{9 \over 2}w^2(2-3w)
(\Omega_m-1)\Omega_m,
\ee
while for the GCG model we find
\be
q_o^{GCG}={3\over2} (1-A_s)-1
\ee
\be
\left.{\d q\over \d z}\right|_0^{GCG}={9\over2}A_s(1-A_s)(1+\alpha),
\ee
\be
\left.{\d j\over \d z}\right|_0^{GCG}= 1+{9 \over 2}\alpha
\left(1-A_s\right)A_s.
\ee

For the redshift range probed by SNe Ia one may neglect terms beyond
the cubic power in redshift in Eq. (\ref{powerDL}). SNe Ia data indicate
that these models have the same deceleration and jerk \shortcite{bertolami2004,bento2005},
that is, they are degenerate for the considered redshift range. Thus, imposing
this equality one finds the relationship between parameters
\be
w=\alpha(A_s-1)-1,
\ee
\be
\Omega_m={(1+\alpha)(1-A_s)\over1+\alpha(1-A_s)}.
\ee

This degeneracy holds for SNe Ia, for the maximum probed redshift of about
$z\approx2$. As the redshift range allowed by GRBs is greater, one hopes to test
higher order terms in Eq. (\ref{powerDL}). At next order, the degeneracy is broken
by the kerk
parameter, that is, even if $q_0^{GCG}=q_0^{XCDM}$, $j_0^{GCG}=j_0^{XCDM}$, one
finds that $k_0^{GCG}\neq k_0^{XCDM}$. For instance, if
one considers $\alpha=1$ and $A_s=0.7$, then $k_0^{GCG}=-0.3$, while
for the corresponding $XCDM$ model (with $w=-1.3$ and $\Omega_m=0.46$),
the kerk is $k_0^{XCDM}=-4.27$.

This procedure can be viewed as a consistency test. Indeed, consider
the GCG model. Once the deceleration, the jerk and the kerk of the Universe
are measured, one can extract the values of
$\alpha$ and $A_s$. These values yield a value for the kerk, call it $k_0^{t}$,
which can be compared to the measured value of the kerk, $k_0^{m}$. Whether
both values
are close or not, one may reach a conclusion regarding the
suitability of the model. The same reasoning can be applied to the $XCDM$ model.
This is, of course, a simplification, since one has to consider the accuracy
of measurements and the statistical significance of each of the parameters
in Eq. (\ref{powerDL}), as well as to consider correlations between them.

%%%%%%%%%%%%%%%%%%%%%%%%%%%%%%%%%%%%%%%%%%%%%%%%%%%%%%%%%%%%%%
%%%%%%%%%%%%%%%%%%%%%%%%%%%%%%%%%%%%%%%%%%%%%%%%%%%%%%%%%%%%%%
\section{Discussion and conclusions.}

We have verified that the medium and high redshift of GRBs do not
impose too strong constraints on the GCG model. At such high redshifts,
the expansion factor for the GCG becomes
\be
H_{ch}(z>>0)=\Omega_{ch}(1-A_s)^{1/(1+\alpha)}(1+z)^3
\ee
where $\Omega_{ch}$ is the CGC density relative to the critical one. Since
we are considering a flat Universe made up mostly by the GCG, $\Omega_{ch}=1$.
Thus, the luminosity distance at such high redshifts depends essentially on
$(1-A_s)^{1/(1+\alpha)}$. It is found  that the confidence
regions for the GCG model approximately follow the line 
$(1-A_s)^{1/(1+\alpha)}=const$. Hence, since $A_s$ and $\alpha$
are strongly correlated by this expression, neither one is constrained.

As for the XCDM model, a high-redshift population
of GRBs is suitable to constrain the total amount of matter, $\Omega_m$, but
are poor probes of the dark energy equation of state. This does not mean that GRBs have no
use for the study of dark energy, or that expanding the Hubble diagram is meaningless.
It is a common trend in the recent literature to argue that a precise prior estimate of $\Omega_m$
enhances the ability of SNe Ia to constrain the dark energy equation of state
\cite{goliath2001,weller2002,pietro2003}. This is also true for GRBs. A high-redshift
sample will constrain $\Omega_m$, while a low-redshift sample
will place constraints on the equation of state. A low-redshift sample alone will fail
to constrain the equation of state since it will be hindered by large uncertainties in 
$\Omega_m$. 
This is somewhat similar to what happens in SNe Ia tests. A low-redshift
sample will constrain the nuisance parameter ${\cal M}$, while the high-redshift
sample constrains the cosmological parameters 
\cite{padmanabhan2003,choudhury2005}.

\begin{figure}
\begin{center}
{\resizebox{0.45\textwidth}{!}{\includegraphics{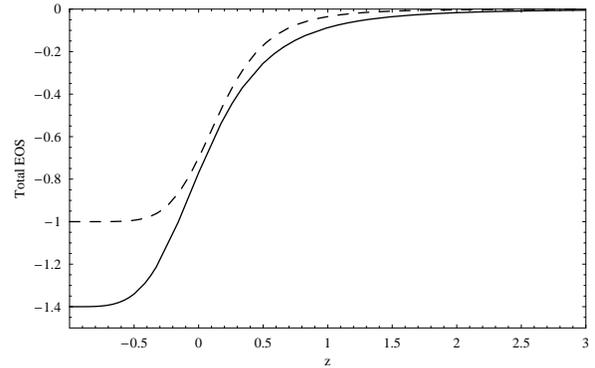}}}
\caption{Total equation of state of the Universe as a function of
redshift. The value $z=-1$ corresponds to the very far future. The dashed
line represents our GCG fiducial model, while the solid line corresponds
to the XCDM phantom model.}
\label{figure8}
\end{center}
\end{figure}

Our study also reveals that the effectiveness of the GRB test depends on the
model under examination. GRBs allow imposing some constraints on the $XCDM$ model, 
but very little can be extracted for the GCG model. This may be explained
by two factors. The first was explained above, and is related to the fact
that the region probed by GRBs depends on an unique function of both parameters. 
The second is related to how late the Universe stops being matter dominated.
As shown in Figure \ref{figure8},
the CGC model stays in the matter dominated phase for a longer
time, thus the period of time during which the dark energy behaviour affects
the cosmic expansion is slightly larger for the XCDM model. Since GRBs 
essentially probe larger redshifts, models in which the transition
between the matter dominated and accelerated expansion is shorter 
cannot be well constrained by this test. By the same reasoning,
GRBs might be actually useful in testing models where the transition
between matter dominated expansion and dark energy driven acceleration
started at larger redshifts.

%%%%%%%%%%%%%%%%%%%%%%%%%%%%%%%%%%%%%%%%%%%%%%%%%%%%%%%%%%%%%%%%%%%%%%%%%%%%%%
%%%%%%%%%%%%%%%%%%%%%%%%%%%%%%%%%%%%%%%%%%%%%%%%%%%%%%%%%%%%%%%%%%%%%%%%%%%%%%

Thus, we find that despite all intrinsic limitations, larger samples of GRBs
can potentially determine the total amount of matter in a XCDM model. This independent
determination of $\Omega_m$ will be useful in SNe Ia studies of the dark energy equation of
state.

Furthermore, GRBs may also be used to break the degeneracy between models, since
at high-redshifts the effect of higher order terms in the Taylor expansion
of $d_L(z)$ must be taken into account.

\bibliography{mnrasmnemonic,astropap}
 
\bibliographystyle{mnras}

\end{document}